\documentclass[sigconf]{acmart}

\AtBeginDocument{%
  \providecommand\BibTeX{{%
    \normalfont B\kern-0.5em{\scshape i\kern-0.25em b}\kern-0.8em\TeX}}}

\copyrightyear{2024}
\acmYear{2024}
\setcopyright{acmlicensed}\acmConference[PACT '24]{International Conference on Parallel Architectures and Compilation Techniques}{October 14--16, 2024}{Long Beach, CA, USA}
\acmBooktitle{International Conference on Parallel Architectures and Compilation Techniques (PACT '24), October 14--16, 2024, Long Beach, CA, USA}
\acmDOI{10.1145/3656019.3676898}
\acmISBN{979-8-4007-0631-8/24/10}

\acmSubmissionID{pact24-p212}

\usepackage[ruled,linesnumbered]{algorithm2e}
\usepackage{algorithmic}
\usepackage{setspace}

\usepackage{multirow}
\usepackage{tablefootnote}
\begin{document}

\title[SZKP: A Scalable Accelerator Architecture for Zero-Knowledge Proofs]{SZKP: A Scalable Accelerator Architecture for \\ Zero-Knowledge Proofs}

\author{Alhad Daftardar}
\email{ajd9396@nyu.edu}
\affiliation{
  \institution{New York University}
  \city{Brooklyn}
  \state{NY}
  \country{USA}
}
\author{Brandon Reagen}
\email{bjr5@nyu.edu}
\affiliation{
  \institution{New York University}
  \city{Brooklyn}
  \state{NY}
  \country{USA}
}
\author{Siddharth Garg}
\email{sg175@nyu.edu}
\affiliation{
  \institution{New York University}
  \city{Brooklyn}
  \state{NY}
  \country{USA}
}

\begin{abstract}
Zero-Knowledge Proofs (ZKPs) are an emergent paradigm in verifiable computing. In the context of applications like cloud computing, ZKPs can be used by a client (called the verifier) to verify the service provider (called the prover) is in fact performing the correct computation based on a public input. A recently prominent variant of ZKPs is zkSNARKs, generating succinct proofs that can be rapidly verified by the end user. However, proof generation itself is very time consuming per transaction. Two key primitives in proof generation are the Number Theoretic Transform (NTT) and Multi-scalar Multiplication (MSM). These primitives are prime candidates for hardware acceleration, and prior works have looked at GPU implementations and custom RTL. However, both algorithms involve complex dataflow patterns -- standard NTTs have irregular memory accesses for butterfly computations from stage to stage, and MSMs using Pippenger's algorithm have data-dependent memory accesses for partial sum calculations. We present SZKP, a scalable accelerator framework that is the first ASIC to accelerate \emph{an entire proof on-chip} by leveraging structured dataflows for both NTTs and MSMs. SZKP achieves conservative full-proof speedups of over 400$\times$, 3$\times$, and 12$\times$ over CPU, ASIC, and GPU implementations.

\end{abstract}



\maketitle

\section{Introduction}
Zero-knowledge proofs (ZKPs) \cite{zkp} are cryptographic protocols that offer an enticing capability: a \emph{prover} is able to convince a \emph{verifier} of the truthfulness of a statement, without revealing any further knowledge of how or why that statement is true. 
For example, one can prove they qualify for a loan \emph{without} revealing bank account details, that one purchased an authentic ticket to enter an event \emph{without} revealing personal details, or that one’s vote was counted \emph{without} revealing which candidate was selected. ZKPs are seeing killer applications in Machine-Learning-as-a-service (MLaaS) \cite{verif_ml} and crypto-currencies \cite{filecoin} (the latter with extensions to electronic voting \cite{vote}). MLaaS providers can use ZKPs to offer services (e.g. recommendation systems) while protecting proprietary model weights that would otherwise be necessary to verify correct operation on clients' inputs. Cryptocurrencies can use ZKPs to enable private transactions while maintaining regulatory compliance \cite{a16zcrypto}.

Like other cryptographic protocols, however, ZKPs incur massive computational overheads. For example, the runtime of a single AES ZKP, i.e., where the prover proves access to the secret key without revealing it, on a CPU is on the order of 100s of milliseconds. This is because of how programs (that are to be verified) have to be transformed into specific linear forms to even construct the proof in the first place, followed by intense computational demands of the underlying cryptographic algorithms.
The impractically high runtimes of ZKPs have motivated research on both GPU and custom hardware acceleration~\cite{pipezk, gzkp, cuzk}. Recent works accelerate two of the most computationally demanding kernels of a ZKP: the number theoretic transform (NTT), and the multi-scalar multiplication (MSM). NTTs are analogs of fast Fourier transforms (FFTs) over finite fields. MSMs compute a weighted sum of points on an elliptic curve; since multiplications in elliptical curve cryptography (ECC) are expensive, they leverage Pippenger's algorithm \cite{pippenger} (see Section \ref{sec:pippenger}) to restructure computation to reduce the number of ECC multiplications in favor of ECC additions.

Yet, prior work has stopped short of fully evaluating the benefits of hardware acceleration for ZKPs. First, these works choose to leave key parts of the ZKP algorithm, for instance, so-called ``Sparse'' MSMs, to software. However, these functions are by no means inexpensive or easy to accelerate. Prior work reveals that Sparse MSMs in software are nearly five times slower than the hardware accelerated components of the ZKP algorithm~\cite{pipezk}. Second, prior work has presented single design point solutions, leaving a large design space of area-performance trade-offs unexplored. This applies as well to the individual MSM and NTT modules; some of these implementations have scalability bottlenecks, requiring, for example, on-chip buffers with a large number of read/write ports and/or complex control logic. And finally, prior work has stopped short of ``full-chip'' simulations/evaluations, leaving, for example, key questions around off-chip bandwidth unaddressed.

SZKP addresses each of these limitations. First, we accelerate \emph{all} steps in online ZKP proof generation. In the process, we uncover interesting and previously unexplored design choices in integrating modules such as Sparse MSMs on the chip. Second, we perform a detailed evaluation of the ZKP accelerator design space, enabled by new scalable designs for MSM and NTT modules with simple control logic and high-level synthesis (HLS) friendly implementations. Full-chip simulations reveal a rich design space of accelerators from large, highly-parallel 800 mm$^{2}$ chips to tiny, sub-50 mm$^{2}$ cores, yielding designs that achieve $12-86\times$ speedups over GPU works while using $50\%$ less area and $3-12\times$ speedups over existing ASICs. 

\section{Background}

\begin{figure*}[t!]
  \centering\
  \hspace{2mm}
  \includegraphics[width=1.02\textwidth]{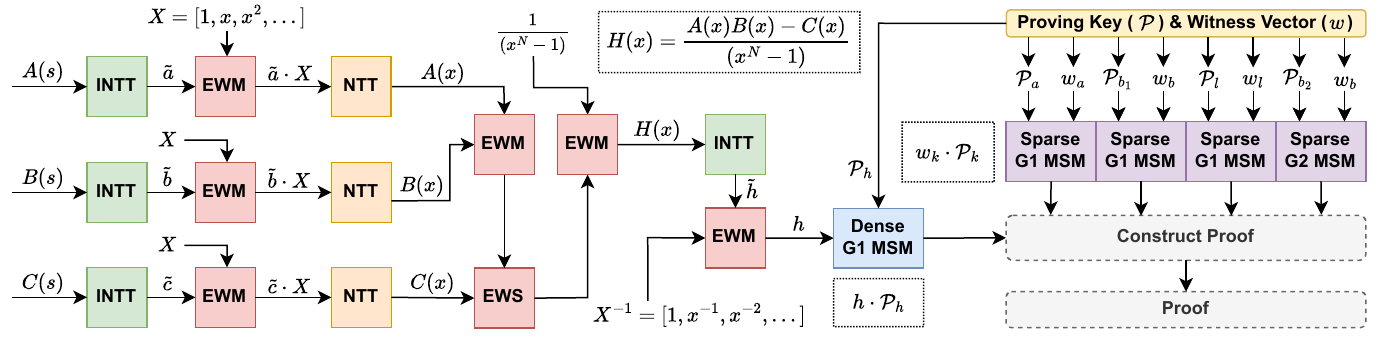}
  \vspace{-0.2in}
  \caption{Groth16 dataflow. This protocol involves 7 (I)NTTs and 5 MSMs to construct a lightweight proof for the verifier. In our design, software provides us $A(s), B(s),$ and $C(s)$, as well as the witness vector $w$ and ECC points derived from the offline-generated proving key.}
  \label{fig:groth-dataflow}
\end{figure*}

\subsection{Zero Knowledge Proofs}
Zero-Knowledge Proofs (ZKPs) allow an entity called the \textit{prover} to demonstrate knowledge of a computation $y = f(w,x)$, on a private witness $w$ and public input $x$ without revealing \emph{anything} about the witness $w$. For instance, a prover could prove knowledge of a secret key that encrypts public input $x$ to output $y$, without revealing the key itself. This ZKP, for example, would allow the prover to access a password-protected cloud service without revealing the password to the cloud. 
Several ZKP protocols have been proposed in literature with different properties. A prominent state-of-art protocol is a zkSNARK, or a \textit{zero-knowledge Succinct Non-interactive ARgument of Knowledge} \cite{zksnark_citation, groth}. zkSNARKs have three main properties: (i) zero-knowledge, meaning the proof reveals no information about the secret witness $w$; (ii) succinctness, meaning the proof is very small, on the order of 100s of bytes, and (iii) non-interactiveness, meaning only one round of communication is performed where the prover sends the proof to the verifier (as opposed to a protocol where multiple rounds of communication are performed). 
Given their wide applicability and usage, this paper focuses on hardware accelerators for zkSNARK proof generation.

\subsection{zkSNARK Protocol Description}
Groth16~\cite{groth}, shown in Figure \ref{fig:groth-dataflow}, is a state-of-art zkSNARK protocol. In this protocol, the prover first generates two keys: (i) a proving key, which is used subsequently to construct the actual proof, and (ii) a verification key, sent to the verifier. This step is time-consuming but is performed offline and only once. 
Then, for each new input $x$, the prover has to construct a proof \emph{online}. This online step is also computationally expensive, hence a potential target for hardware acceleration~\cite{gzkp, pipezk, cuzk}. 
Proof generation in Groth16 involves two basic operations: MSMs and NTTs. These are described next, followed by the overall dataflow of Groth16.

\subsubsection{MSMs}\label{msm_intro}
MSMs compute dot products, i.e., 
$\sum_{i=0}^{n-1} a_{i}P_{i}$
where $P_{i}$ are elements in a cyclic group $G$, typically, 3-dimensional points on an elliptical curve,  
and $a_i$ are scalar integers. The individual scalars and coordinates for points can range from 254-753 bits wide depending on the elliptical curve \cite{libsnark}. Compared to a typical dot product operation, multiplication between a scalar and a point, e.g. $a_{i}P_{i}$, is computationally expensive and achieved via repeated additions of point $P_{i}$. 
Point additions are themselves costly; depending upon the elliptical curve, a point addition consists of anywhere between 16 and 80 modular multiplications alone. 

Groth16 computes two types of MSMs: Sparse and Dense. Sparse MSMs differ from Dense MSMs primarily in the values of the scalars. As the name suggests, Sparse MSM scalars are typically $0$ or $1$, and usually account for $90\%$-$99\%$ of all scalars. This is because scalar values come from intermediate outputs of program execution, and $0$s and $1$s relate to the outcomes of branches, conditionals, and other nonlinearities in the program execution that is being verified.
This means that roughly half the points can be ignored and half can be directly added once, while the remaining 1-10\% of points require more complex processing. In contrast, scalars for Dense MSMs are typically uniformly distributed \cite{libsnark}.

In the Groth16 protocol, there are four Sparse MSMs. Three of these are performed on points belonging to an elliptic curve group G1, as is the Dense MSM. The fourth MSM is performed on points from an elliptic curve group G2. The latter operation is distinct in that it involves computations on pairs of points within G2, which is necessary for the pairing-based operations that underpin the Groth16 proofs. Consequently, the G2 MSM is computationally more expensive than G1. Furthermore, one of the Sparse G1 MSMs and the Sparse G2 MSM also exhibit sparsity in their \textit{points}. In ECC, this manifests as the ``point at infinity'' $\mathcal{O}$, which essentially has the same properties as the number 0 in normal addition.

\begin{figure}
  \hspace{-5mm}
  \centering
  \includegraphics[width=0.5\textwidth]{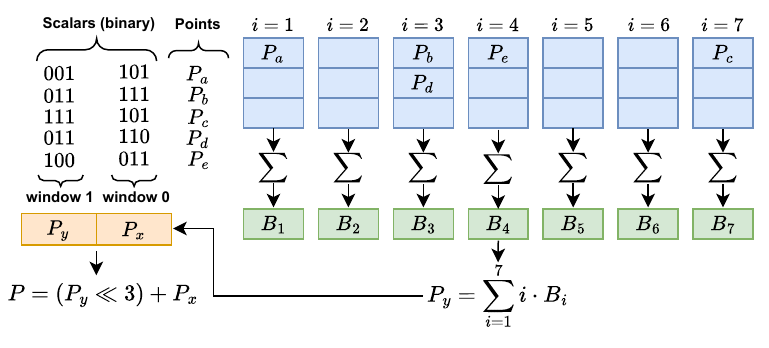}
  \vspace{-0.3in}
  \caption{Pippenger's algorithm. This example demonstrates computation of window 1, with $P_x$ already computed for window 0. The final result involves doubling $P_y$ 3 times before adding it with $P_x$}
  \label{fig:buckets}
  \vspace{-3mm}
\end{figure}

\subsubsection{NTTs and Polynomial Computation}
The Number Theoretic Transform is an analog of the Fast Fourier Transform with elements that lie in a field, for instance, integers modulo a prime $p$. NTTs are used for performing polynomial multiplications, which are equivalent to the convolution of polynomial coefficients. Instead, it is faster to compute the NTT of the coefficients of each polynomial, element-wise multiply (EWM) the NTT outputs, and perform an inverse NTT (INTT).  Groth16 involves multiplications, divisions and subtractions over input polynomials implemented using a sequence of NTTs, EWMs and INTTs.  In hardware, typically the Cooley-Tukey NTT is used, based on the Cooley-Tukey FFT \cite{cooley}. Several prior works \cite{f1, clake, bts, ark, sharp, gzkp, pipezk, rpu} propose techniques to accelerate the NTT, including two prior works on ZKPs.

\subsubsection{Overall Dataflow of Groth16} While we cannot do full justice to Groth16 here, we will highlight some key concepts relevant to hardware acceleration. Groth16, as well as several other zkSNARK protocols, represents the computation $f$ as a rank-1 constrained system (R1CS) that represents the inputs and outputs of each intermediate computation in $f$ via three vectors, $a$, $b$ and $c$. These vectors are transformed into polynomials $A(x)$, $B(x)$ and $C(x)$, using which Groth16 first computes:
\begin{equation*}
    H(x) = \frac{A(x)B(x) - C(x)}{x^N - 1}
\end{equation*}
where $N$ is the NTT length. All three sequences start in the NTT domain, and each undergoes an INTT, an EWM, and then another NTT. Then, $H(x)$ is computed, with the division operation performed as a multiplication by a scalar. $H(x)$ is still in the NTT domain, so it is processed with an INTT and an element-wise multiply to yield the set of scalars $h$. 

The elements of $h$ form the scalars that are used to compute a Dense MSM with points derived from the proving key (computed offline). In parallel, the proving key and witness vectors are also used to compute the four sparse MSMs. Together, the MSM outputs are used to construct the final proof sent to the verifier.

\section{The SZKP Architecture}

\subsection{Dense MSM Architecture}
SZKP's Dense MSM architecture uses Pippenger's algorithm~\cite{pippenger}, commonly deployed in a range of ZKP implementations~\cite{cuzk, pipezk, gzkp}. We review Pippenger's before describing our architecture. 

\subsubsection{Pippenger's algorithm for MSMs}
\label{sec:pippenger}
MSM computations are prohibitively expensive for high bit-width scalars because of the potentially large number of point additions needed to compute each term in the MSM. The commonly used alternative is Pippenger's algorithm (see Figure \ref{fig:buckets}), where $\lambda$-bit scalars are broken up into narrower $W$-bit \textit{windows}. For instance, a 256-bit scalar might be broken up into 64 4-bit windows.  

Then, each scalar within a window maps to one of $B=2^W-1$ buckets, omitting the $0^{th}$ bucket. For example, for 5-bit windows, there are 31 buckets. Points are accumulated into buckets corresponding to their 5-bit scalar values---equivalently, all points corresponding to the same 5-bit scalar are mapped to the same bucket. These points are then added together and multiplied with their corresponding scalar value at the very end. Finally, the results of each bucket are summed to yield a final value for the window. 

Once all windows complete their bucket summations, a global reduction is performed by ``bit-shifting'' each $i^{th}$ window's reduced sum $b_i$ times, where $b_i$ is the offset of window $i$ in the scalar. For this operation, point doubling is used in lieu of point addition, i.e., without incurring point-equality checks inherent to point additions.

Pippenger's algorithm offers two opportunities for parallelism: (1) points in each bucket within a window can be simultaneously added; and (2) multiplications across windows are fully independent until the final global reduction. This implies that two different windows can compute their point additions entirely in parallel. However, both approaches assume unrestricted and content-based access to scalars and points which is challenging in practice. 

Prior works have exploited these opportunities in both GPU and ASIC. However, they encounter heavy preprocessing overheads \cite{gzkp} or are not scalable because of how they handle data on-chip and because they offload Sparse MSM compute to CPU \cite{pipezk}.

\begin{figure*}[t!]
  \centering
  \includegraphics[width=\textwidth]{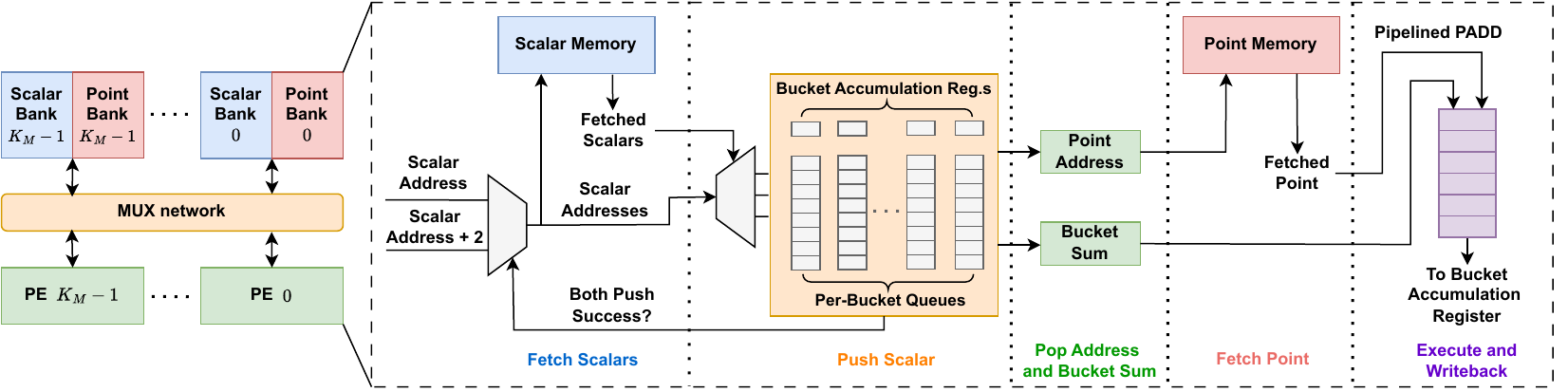}
  \vspace{-0.2in}
  \caption{Pipeline architecture for a single Dense MSM. Buckets store point addresses. Queue selection policy can be RR, Max-\textit{r}, or LQ. Writebacks always succeed. Each PE reads from one bank of scalars and points at a time, avoiding memory contention.}
  \label{fig:msm_pipe}
\end{figure*}

SZKP relies on a PE-array to exploit the high degree of parallelism inherent in Pippenger's algorithm. Each PE handles the computation for one $W$-bit window of scalars at a time, reading scalars and points from pre-loaded on-chip scalar and point buffers. We first describe the architecture of a single PE, and then how SZKP handles multiple PEs on the chip.

\subsubsection{Single PE Design}
Figure~\ref{fig:msm_pipe} shows a pipeline diagram of a single PE. As noted, in Pippenger's algorithm, points are fetched into $B=2^W-1$ buckets and each bucket independently accumulates points mapped to it. In theory, therefore, this work can be parallelized using $B$ parallel point adders (PADDs). Unfortunately, point addition is \emph{slow}; consistent with prior work, the PADDs we synthesized have latencies in the tens of clock cycles (see Table~\ref{tab:ii}) and dominate chip area. Therefore, we allocate a single, fully-pipelined PADD to each PE. Prior work does the same but requires complex control logic to maximize PADD utilization. We show a much simpler scheme that can achieve equal or better utilization.

\textbf{Storing Point Addresses:} Each MSM PE has $B=2^W-1$ buckets, each with a queue of depth $D$ implemented as a FIFO. Each bucket also has a \emph{single} bucket accumulation register, which stores partially accumulated values of points mapped to that bucket. In each cycle, 2 scalars are fetched from the scalar memory to index the buckets, but instead of pushing \textit{points}, we push \textit{addresses} into the respective buckets' queues. Points are typically large, ranging from 762-2259  bits in total, depending upon the elliptical curve, but addresses to a typical 16K point buffer are only 14 bits. Thus, we can provision deep queues, thus ensuring the PADD always has a sufficient supply of operands to compute from and minimizing stalls to the PADD. Additionally, the control logic in the fetch stage is simple because we always fetch consecutive scalars unless we stall because a bucket's queue is full. Prior work \cite{pipezk} has a complex fetch mechanism that must fetch the points themselves into either a bucket register or one of two FIFOs, having to also face contention with writebacks to the bucket register after a sum is computed from the PADD.

\begin{figure}
  \centering
  \hspace{-.5cm}
\includegraphics[width=0.45\textwidth]{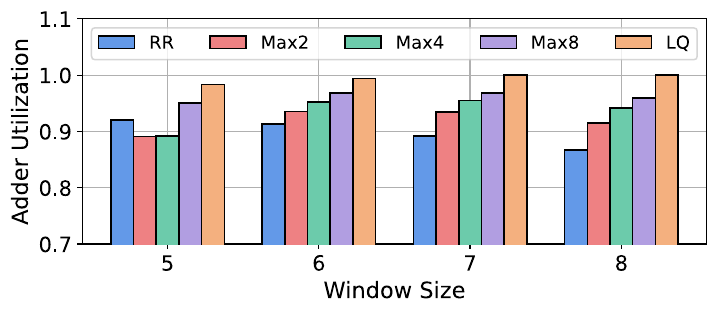}
\vspace{-0.2in}
  \caption{PADD utilization across varying window sizes for round robin (RR), longest queue (LQ) and Max-$r$. Max-8 and LQ consistently have more than 90\% utilization.}
  \label{fig:au}
  \vspace{-3mm}
\end{figure}

\textbf{Round-Robin Scheduling:} How can we ensure the PADD is maximally utilized? We start with the simplest implementation and then iteratively refine it. As scalars are fetched, in parallel, the PADD iterates over buckets in \emph{round-robin} order. If a bucket's queue is non-empty, it pops an address and uses it to fetch the corresponding point from the point buffer and add it to the value in the bucket's accumulation register. If the bucket's queue is empty, it inserts a bubble into the PADD and moves on to the next bucket. Interestingly, we find that as long as the number of buckets, $B$, is greater than the PADD latency, $t_{add}$, even this simple design has more than $85\%$ utilization (see Figure~\ref{fig:au}). The reason is two-fold: (1) by the time the PADD returns to a bucket, the previous add issued from that bucket has completed, ensuring that a new add can be issued unless the bucket's queue is empty; and (2) scalar values in dense MSMs tend to be uniformly distributed; thus  the probability of the bucket's queue being non-empty in $t_{add}>B$ clock cycles is high. We introduce two further optimizations to this architecture. 

\textbf{Longest Queue Policy:} To further increase utilization, we replace round-robin with a longest available queue (LQ) dispatch policy. LQ uses simple tournament-style logic to schedule the next add operation from the deepest buffer with a valid accumulated value. As shown in Figure~\ref{fig:au}, LQ achieves $97-99\%$ utilization for 5- to 8-bit windows. We also implement a simpler version of LQ that dispatches from the largest of $r$ consecutive  buffers (Max-$r$). For example, Max-8 achieves more than $95\%$ utilization as shown in  Figure~\ref{fig:au}. In contrast, prior work \cite{pipezk} feeds the PADD from one of two FIFOs, each holding two full points. While this keeps the PADD fed, they also require a spillover FIFO that store the results of add operations that could not be written back to a bucket register, necessitating further area for storing points. In this scheme, it is trivial to show that if the spillover FIFO is not prioritized for pushing points, then the system deadlocks, and forward progress cannot be made leading to functional correctness not being satisfied.

\subsubsection{Scaling to Multiple PEs} SZKP allocates $K_M$ PEs to exploit parallelism across different scalar windows that operate on the same set of points. In prior work~\cite{pipezk}, each PE reads from a single point buffer shared by all PEs. Unfortunately, since the PEs operate on different scalar windows, they operate out-of-sync  since they stall at different points. 
To enable PEs to read from different locations in the point buffer, prior work assumes a multi-ported point buffer with two ports per PE, presenting a scalability bottleneck.

In SZKP, we partition the scalar and point buffers into $K_M$ banks, each with 2 read/write ports. Each PE operates on a \emph{different} bank; when each PE has completed operating on the points in its bank, it switches to the next bank and so on. Specifically, in round $j$, PE $i$ operates on bank $(i+j)\,\%\,K_M$. We note that this scheme does introduce synchronization delays since each PE waits for others to finish. However, although the exact execution traces of the PEs differ, their completion times differ negligibly because scalar values for Dense MSMs are uniformly distributed and hence each PE has the same stall probability. In other words, the PEs share the same statistical behavior which averages out over sufficiently many points. The logic for each PE switching to a different memory bank is simple and can be handled by a simple crossbar, thereby avoiding memory contention that would otherwise occur for the same set of points. Additionally, MSM sequences are long, so to minimize the number of off-chip accesses, the PEs collectively compute the MSM on \textit{all} windows (storing per-window partial bucket sums in local on-chip memories) before fetching more points and scalars into the point and scalar memories. This enables us to have relatively low MSM bandwidth costs in spite of the large bit-widths.

\begin{algorithm}
    \small
    \SetKwFor{For}{for}{do}{}
    \SetInd{0.5em}{0.8em}
    \caption{$\mathsf{Bucket Reduction}$}
    \label{alg:bucket_reduce}
    \DontPrintSemicolon
    
    $s_r \leftarrow \mathcal{O}\,\,\,$\tcp{running sum}
    $s_t \leftarrow \mathcal{O}\,\,\,$\tcp{total running sum}
    \For{$i=2^W-1$ \KwTo $1; \ i = i - 1$}{    
        $s_r \leftarrow s_r + B_i$\;
        $s_t \leftarrow s_t + s_r$\;
    }
    \Return $s_t$
\end{algorithm}
\vspace{-3mm}

\textbf{Bucket Reduction}:
After a PE processes all points for a window, the window must compute a bucket reduction, $ \sum_{i=1}^{2^W - 1} i\cdot B_i$, where $B_i$ is the accumulated value in bucket $i$. Prior work \cite{pipezk} performs bucket (and window) reductions in software because these typically constitute less than 0.1\% of the CPU runtime. However, with the rest of the MSM computation accelerated on-chip, reductions become a larger portion of the runtime if offloaded to CPU. In SZKP, we observe that the recursive Algorithm \ref{alg:bucket_reduce} from \cite{bucket_reduce_libsnark_citation} admits a hardware-friendly implementation with simple control logic. This algorithm iteratively computes a running sum in one pass with two serial additions per bucket, with a total latency of $(2t_{add})(2^W - 1)$ cycles.

\textbf{Window Reduction:}
\label{sec:window_reduction}
The final step is to take the sum in each window and compute its offset based on its location in the overall scalar. This is analogous to bit-shifting with integers, but because we are computing on elliptical curve points, we instead perform point-doubling. Note that the most-significant bits (in the most-significant window) have to be doubled the most times. For example, in a 254-bit scalar with 5-bit windows, the highest window has to be scaled by roughly $2^{250}$, meaning 250 \textit{serial} point doubling operations. To account for this, we adopt the approach in the CPU implementation \cite{libsnark} where we start computing the accumulated sum with the MSB windows and iteratively compute point doubling towards the LSB windows. With multiple PEs, each PE can handle the doubling responsibilities for 1 window. With $K_M$ PEs, $W$-bit windows, the latency of window reduction is roughly
\begin{equation*}
    (K_M \cdot W \cdot t_{dbl} + t_{add})\lambda / W
\end{equation*}
The critical path for each group of $K_M$ PEs is the number of point-doubling operations needed by the PE handling the accumulated result (from the prior group of windows). The maximum number of point doubles done in a group is $W \cdot K_M$ (the bit-width spanned by all PEs). Since the LSB window doesn't need to be doubled within a group of PEs, all PADD units (which support doubling) are utilized. 
\vspace{-2mm}

\subsection{Supporting Sparse MSMs}

For the majority of the Sparse MSM computation, Sparse MSMs can be viewed as Dense MSMs with only one bucket that corresponds to a scalar value of 1. The Dense MSM architecture loads a point from point memory and adds it to an accumulated value. Instead, in a Sparse MSM core, we first load all points with a scalar of 1 into the point memory, fetch two points at a time, feed them to the PADD, and write back the result to point memory, resulting in near perfect utilization. This process continues until the desired sum is computed. Afterwards, we compute the scalar multiplications for the remaining 1\% of scalars using the same Pippenger approach as for Dense MSMs. While prior work implements Sparse G1 MSMs (without any reductions), they offload Sparse G2 MSMs (and all G1 reductions) to CPU \cite{pipezk}; in contrast, SZKP accelerates Sparse G2 MSMs with a dedicated G2 MSM core.

\subsubsection{Separate Vs. Shared Hardware} \label{sh_v_sp} Since both Sparse and Dense MSMs in group G1 use the same PADDs, we explore PADD sharing between the two operations. We note in the overall Groth16 dataflow that the Sparse and Dense MSMs operate in parallel; thus, sharing hardware does cause an artificial dependency between these operations that can increase overall latency. On the other hand, separate hardware for Sparse and Dense MSMs increases chip area. We compare these alternatives in our empirical evaluations.

\subsection{Sparse G2 MSM Optimizations}
A fully pipelined G1 adder by itself is large, requiring 16+ modular multipliers. But a G2 adder is much larger, requiring 80+ modular multipliers. Because Sparse G2 MSMs are executed independently of other kernels and exhibit sparsity in their \textit{points}, they are not on the critical path. We can significantly reduce modular multiplier counts of Sparse G2 MSMs by pipelining them less. We investigate this by building G2 PADD units with initiation intervals (II) from II=1 to II=4\footnote{In HLS terminology, II represents a module's throughput; smaller IIs represent more deeply pipelined modules.}. We also implement this for G1 PADD units and tabulate the multiplier counts in Table \ref{tab:ii}.

\begin{table}[ht]
\centering
\caption{Modular multiplier counts for G1 and G2 PADDs for different initiation intervals (IIs) and ECC bit-widths.}
\label{tab:ii}
\begin{tabular}{|c|c|c|c|c|c|}
\hline
\textbf{PADD} & \textbf{Cycles} & \textbf{II=1} & \textbf{II=2} & \textbf{II=3} & \textbf{II=4}\\ \hline  \hline
\textbf{G1 (254b)} & 30  & 16 & 8 & 7 & 5  \\ \hline
\textbf{G2 (254b)} & 55  & 80 & 43 & 36 & 22  \\ \hline
\textbf{G1 (753b)} & 38  & 23 & 15 & 9 & 7  \\ \hline
\textbf{G2 (753b)} & 67  & 85 & 45 & 38 & 25  \\ \hline
\end{tabular}
\end{table}

\begin{figure*}
  \centering
  \includegraphics[width=0.98\textwidth]{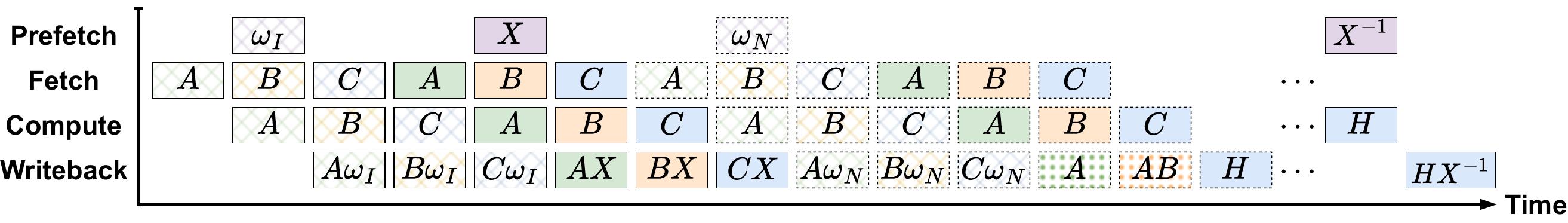}
  \vspace{-3mm}
  \caption{Example Polynomial Computation Pipeline Schedule. This simple schedule assumes an $M \times M$ matrix and $M$ NTT PEs. Slots with solid borders represent an INTT phase, while dashed borders represent an NTT phase. Slots with cross hatches represent column-wise operations, while slots with solid fill represent row-wise operations. Slots with dots represent values that used operands stored on-chip instead of being prefetched. $\omega_I$ are INTT twiddles, $\omega_N$ are NTT twiddles, and X are generators.}
  \label{fig:poly_sched}
\end{figure*}

\begin{figure}
  \centering
  \includegraphics[width=0.5\textwidth]{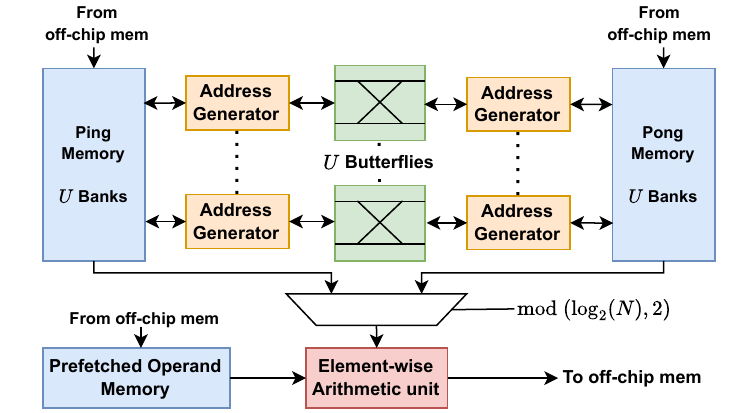}
  \vspace{-0.2in}
  \caption{Single NTT PE with simple, static control logic for address generation}
  \label{fig:nttarch}
  \vspace{-5mm}
\end{figure}

\vspace{-2mm}
\subsection{NTT Architecture}
\textbf{Four-Step NTT}: In Groth16 protocols, the NTT sequences are long, ranging from $2^{14}-2^{20}$ and potentially longer sequences. Because large NTT datapaths are prohibitive in cost, we use the Four-Step algorithm \cite{fourstep} used by prior works \cite{f1, clake, ark, pipezk} to decompose an $N$-point NTT into a series of $\sqrt{N}$  $\sqrt{N}$-point computations. Thus, we only need to construct an NTT core that computes on up to $2^{10}$ points to support power-of-2 NTT evaluations up to $2^{20}$-point NTTs. Each Four-Step (I)NTT involves transforming the input polynomial into a two-dimensional array, computing the (I)NTT first on the columns, scaling by twiddle factors (different from those used within the (I)NTT), tranposing the matrix, and then performing (I)NTT on the rows. For each of these steps, we fetch inputs from off-chip memory, compute on our NTT core, and write back results to off-chip memory in either a transposed or normal manner, depending on the step being computed.

\textbf{Constant-Geometry Topology}: Most prior NTT accelerators are based on the Cooley-Tukey algorithm ~\cite{cooley}. However, Cooley-Tukey exhibits irregular data flow from stage to stage, resulting in irregular memory access patterns~\cite{ncsu, ntt_ss_roy}. 
For this reason, HLS tools struggle to synthesize high-performance memory-based Cooley-Tukey NTT implementations. Instead, we find that a constant-geometry implementation based on the Pease and Korn-Lambiotte algorithms ~\cite{pease, kl} synthesizes high-utilization NTT hardware and much faster than Cooley-Tukey implementations.

In constant-geometry NTTs, the read and write addresses for each butterfly unit in a stage remain fixed, allowing for simple muxing between butterflies and memory buffers. We implement our NTT network using $U$ parallel butterflies and a double-buffered, dual-ported ping-pong on-chip memory with $U$ banks. True dual-porting enables full pipelining of reads from ping (pong) buffers, butterfly compute, and writes to pong (ping) buffers in each stage. Our constant-geometry NTT implementations are competitive with state-of-art hand-designed NTT accelerators~\cite{f1, bts, pipezk, ncsu}. 

Note that manually-coded Cooley-Tukey NTTs (using a memory-based architecture as we do) would be equally performant to our HLS constant-geometry NTT because the irregular memory accesses per stage could be managed by a dedicated controller, thereby allowing high butterfly utilization. However, the logic for this is complex (adding area overheads) and requires special care to avoid memory conflicts as we scale the number of butterflies \cite{ntt_ss_roy}. Manually-coded NTTs using pipelined architectures \cite{garrido, pipezk} for large lengths require dedicated SRAM delay buffers (and associated controllers) for each stage. These too add non-trivial overheads in area and interconnect. $N$-point constant-geometry NTTs have simpler control logic, and in turn, are faster to design and prototype with HLS tools. However, they are out-of-place and require $2N$ words for storing intermediate computations. As we show later, MSMs are typically area-dominant in SZKP, so the out-of-place memory costs are worth the savings in design complexity.

\textbf{Operands}: 
In the Four-Step (I)NTT, the output of column-wise (I)NTTs are multiplied by twiddle factors before being written back to memory. INTT outputs are scaled by a series of multiplicative generators on the finite field (the vector $X$ in Figure \ref{fig:groth-dataflow}), $B(x)$ is multiplied with $A(x)$, $C(x)$ is subtracted from $A(x)B(x)$ and $H(x)$ is scaled by inverse generators ($X^{-1}$) on the finite field. For post-(I)NTT scaling/subtraction, we call these twiddle factors, generators, and intermediates \textit{operands}. Operands are the same length as the (I)NTT, and we prefetch them from off-chip memory while (I)NTTs are computed by the butterflies. We then pipeline the element-wise operand arithmetic with writeback to off-chip memory.

We can significantly reduce the prefetch bandwidth for the twiddle factors and generators because they can be computed on-the-fly given initial columns of the twiddle and generator matrices. Then, by interleaving the execution of the $A$, $B$, and $C$ (I)NTTs, we maximize operand reuse and generate them on-the-fly in our element-wise arithmetic unit, eliminating most of the prefetch bandwidth and masking most of the matrix transpose latencies. The operands that cannot be generated on-the-fly are $A(x)$ for multiplication with $B(x)$ and $A(x)B(x)$ from which $C(x)$ is subtracted; these operands can directly be loaded into the operand memory buffer instead of writing back to off-chip memory.

Finally, the four-step (I)NTT's column/row-wise operations can be performed independently prior to the matrix transpose. This allows for a multi-PE (I)NTT where each PE handles one column/row (I)NTT without the need for complex interconnect for transferring data. This enables easy scaling of the number of PEs.

Figure \ref{fig:poly_sched} shows a simplified example of the polynomial computation schedule to highlight the kernel-level pipelining, with each on-chip compute and fetch, prefetch, and writeback from off-chip memory (except where noted) running concurrently. For simplicity, we display the same matrix names ($A, B, C, \text{or } H$) for all intermediate and final results. Here, we can see how interleaved scheduling allows us to complete all column-wise computations before reading in data row-wise. This helps mask serial matrix transpose latencies.

\textbf{Bit-reversals}: Since our NTT unit supports both NTT and INTT, it needs to handle bit-reversals. Our NTT does this by utilizing the chaining property as noted by prior designs \cite{pipezk, no_bitrev} to avoid bit-reversals. However, if we assume the input polynomials are in natural input order, the final INTT step will output scalars in bit-reverse order. Because the points used by the Dense MSM are generated by the proving key during one-time key generation, we assume the key generation software bit-reverses these points for us. Since point addition is commutative, the order in which we compute the MSM does not matter, thus, elimination of bit-reversal overhead still maintains functional correctness.

\section{Evaluation Methodology}
\begin{figure*}
  \centering
  \hspace{10mm}
  \includegraphics[width=0.9\textwidth]{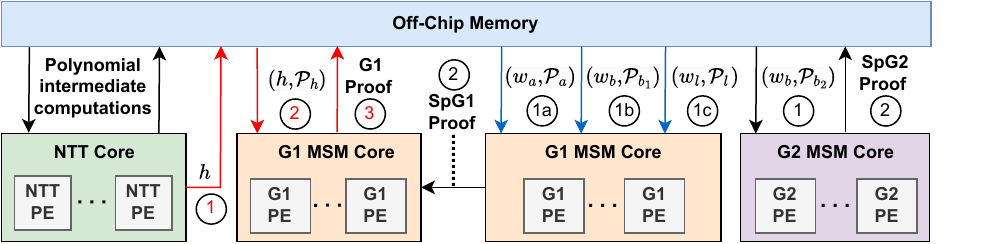}
  \vspace{-.1in}
  \caption{SZKP Chip Architecture. Numbers indicate order of operations. Red arrows and numbers indicate the critical path from polynomial computation through the Dense MSM. Blue arrows indicate serialization of the 3 Sparse MSMs through the G1 core dedicated for Sparse MSMs. The final proof is constructed in two parts, a G1 component and a G2 component, that get written back to off-chip memory}
  \label{fig:chip_arch}
  \vspace{-2mm}
\end{figure*}

SZKP is, as far as we know, the first ASIC to accelerate entire zkSNARK proofs on-chip, including Sparse G1/G2 MSMs and MSM reductions. Figure \ref{fig:chip_arch} details the full SZKP architecture. As part of our evaluation, we conduct a comprehensive design space exploration of both individual modules and the full chip. We additionally investigate resource sharing among G1 MSMs as well as the effect of off-chip bandwidth constraints on total proof-generation time.
\vspace{-2mm}

\subsection{Performance Modeling}
We used Catapult HLS 2022 to generate the RTL for pipelined field adders and multipliers (we used Montgomery multipliers \cite{mont} as in prior work \cite{f1, pipezk}), constant-geometry NTT PEs, and G1 and G2 PADD units with different IIs to study area vs. runtime tradeoffs. 
We synthesized RTL using Synopsys DC Compiler with a TSMC 22nm technology library and a Synopsys 22nm Memory Compiler for area estimation with a 300 MHz clock to match prior ASICs \cite{pipezk}. We also separately synthesized the longest queue (LQ) dispatch policy and confirmed that it clocks at 1 GHz, much faster than the 300 MHz target. 

We used HLS-generated modules to determine PADD and NTT latencies for performance modeling. We developed a cycle-accurate simulator to model the runtime of the Dense MSM pipeline using the LQ dispatch policy. We used this simulator along with analytical models for Sparse MSM computations and bucket/window reductions. For polynomial computation, we used single-NTT latencies to model the total latency and constructed traces of the off-chip memory accesses for inputs, operands, and outputs. We initially assumed unconstrained memory bandwidth to model peak theoretical performance, and later imposed bandwidth constraints for a range of memory technologies. We also constructed power traces to model thermal design power and power density.
\vspace{-1mm}

\subsection{Benchmarks}
\label{sec:bench}
We evaluate our architecture using workloads from Jsnark \cite{jsnark}, a Java based library with Libsnark \cite{libsnark} as its backend. Libsnark is a state-of-the-art CPU implementation of Groth16, supporting several elliptic curves used for zkSNARK computations.

We focus our analysis on the BN128 and MNT4753 curves which both provide 128 bits of security and are commonly used in prior work \cite{cuzk, pipezk, gzkp}.  The former has an underlying bit-width of 254 for scalars and points, while the latter uses 753 bits. We compare SZKP with CPU using BN128, with a prior ASIC using BN128 and MNT4753, and with prior GPUs using MNT4753. This is done to reflect the range of areas we typically see as SZKP is scaled to higher bit-widths. We first examine the performance of individual modules and then use Jsnark workloads to measure the time to compute full proofs against CPU, ASIC, and GPU.

\vspace{-1mm}
\begin{table}[ht]
\centering
\caption{Design Space of SZKP Architecture. We evaluate all combinations of these design knobs.}
\label{tab:design_space}
\vspace{-3mm}
\begin{tabular}{|c|c|c|}
\hline
\textbf{Module} & \textbf{Design Parameter}            & \textbf{Values}   \\ \hline \hline
\textbf{MSM} & PEs ($K_{M}$)               & 1, 2, 4, 8, 16    \\ \hline
\textbf{MSM} & Window Size ($W$)       & 5, 6, 7, 8    \\ \hline
\textbf{MSM} & Points/Window ($PPW$) & 1K, 2K, 4K, 8K, 16K \\ \hline
\textbf{MSM} & PADD Pipelining (II)              & 1, 2, 3, 4    \\ \hline
\textbf{NTT} & PEs ($K_{N}$) & 1, 2, 4, 8 \\ \hline
\textbf{NTT} & Butterflies ($U$) & 1, 2, 4, 8, 16, 32 \\ \hline
\end{tabular}
\end{table}

\section{SZKP Evaluation}
We now evaluate the area and performance of a range of different SZKP accelerators using the design space exploration parameters listed in Table \ref{tab:design_space}. In our evaluations, we begin by reporting the area and power of individual components. We then use synthetic data to determine runtime and speedup of SZKP's Dense MSMs and NTTs over CPU, ASIC, and GPU designs.

Next, we report full-chip results. Since prior work did not accelerate Sparse MSMs in hardware, we highlight the impact of design choices including shared vs. separate hardware for G1 MSMs and the benefits of optimizing II for Sparse MSM PADDs. We then identify and analyze several representative zkSNARK designs at different area-performance points, show the impact of bandwidth on accelerator performance, and report full-chip speedups relative to CPU, ASIC, and GPU.

\begin{table}[ht]
\centering
\caption{Area and Power of Components}
\label{tab:componentwise_area_power}
\vspace{-2mm}
\begin{tabular}{|c|c|c|c|c|}
\hline
\multirow{2}{*}{\textbf{Component}} & \multicolumn{2}{c|}{\textbf{Area (mm$^2$)}} & \multicolumn{2}{c|}{\textbf{Power (W)}} \\ \cline{2-5}

 & \textbf{254b} & \textbf{753b} & \textbf{254b} & \textbf{753b} \\ \hline \hline
\textbf{Modmul}         & 0.30 & 2.89  & 0.12 & 0.78   \\ \hline
\textbf{NTT Butterfly}  & 0.31 & 2.90  & 0.13 & 0.81   \\ \hline
\textbf{G1 PADD II = 1} & 5.05 & 67.33 & 2.14 & 18.50 \\ \hline
\textbf{G1 PADD II = 2} & 2.56 & 43.91 & 1.08 & 12.07 \\ \hline
\textbf{G1 PADD II = 3} & 2.52 & 26.35 & 1.06 & 7.24  \\ \hline
\textbf{G1 PADD II = 4} & 1.61 & 20.49 & 0.67 & 5.63  \\ \hline
\textbf{G2 PADD II = 4} & 8.72 & 73.74 & 2.72 & 20.20 \\ \hline
\end{tabular}
\vspace{-2mm}
\end{table}

\subsection{Individual MSM and NTT Unit Evaluations}
\begin{figure}
  \centering
  \hspace{-5mm}
  \includegraphics[width=0.5\textwidth]{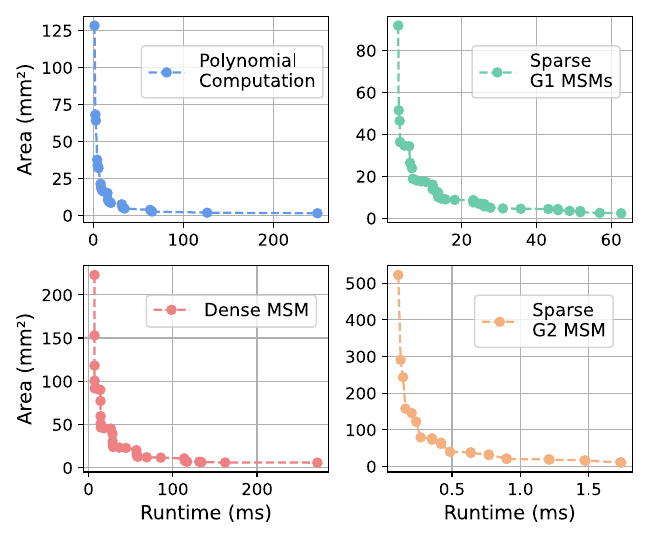}
  \vspace{-0.3in}
  \caption{Pareto Curves for individual SZKP modules. Sparse G2 MSMs are the largest module by area but contribute to less than 0.5\% of total proof generation time}
  \label{fig:combined_pareto}
  \vspace{-3mm}
\end{figure}

Table \ref{tab:componentwise_area_power} shows the area and power for different components synthesized via HLS. For 254b, HLS generates 30-cycle G1 PADDs and 55-cycle G2 PADDs; for 753b, HLS generates 38-cycle G1 PADDs and 67-cycle G2 PADDs (see Table \ref{tab:ii}). We estimate power using pessimistic 50\% switching activity. Figure ~\ref{fig:combined_pareto} plots area vs. runtime Pareto plots for each individual module running the Auction workload on 254b datatypes. We note that although Sparse G2 MSMs consume the most area, they are much faster than and can be computed in parallel with other computations. Thus, in practice, full-chip Pareto designs only instantiate the smallest Sparse G2 units which are still well off the critical path. Of the remaining modules, Dense MSMs consume the most area, followed by NTTs and Sparse G1 MSMs.

\subsubsection{SZKP vs. CPU}
Table \ref{tab:combined_speedup} details the speedup of our Dense MSM in isolation and NTT in isolation versus CPU. For fair comparison, we pick an MSM and NTT design point that is roughly iso-area to a single core on our CPU, an AMD EPYC 7502 32-core processor running either a single Dense MSM or single polynomial computation. The CPU processor consists of four 8-core chiplets, each of die size 74 mm$^2$ fabricated in 7nm, with a memory bandwidth of 204.8 GB/s \cite{amd_specs, tech_powerup, amd_article}. We estimate the core area to be 9.25 mm$^2$, and using a scale factor of 3.6 based on prior work \cite{haac, 28nmto16nm}, we pick designs close to 30-33 mm$^2$ in 22nm and estimate a speedup factor of 1.7 to model scaling down to 7nm \cite{28nmto16nm}. Under these constraints, we choose an MSM with 4 PEs, 8-bit windows, and 4096 points per window, and an NTT unit with 4 PEs and 16 butterflies/PE.

\subsubsection{SZKP vs. PipeZK}

We compare individual module speedups with PipeZK \cite{pipezk}, the only known ASIC that accelerates full zkSNARK proofs, on 254b data in Table \ref{tab:combined_speedup_pipezk}. For this, we pick an MSM with 4 PEs, 8-bit windows, and 2048 points per window (26 mm$^2$ vs. PipeZK's 35 mm$^2$ MSM), and an NTT with 4 PEs and 8 butterflies/PE (18 mm$^2$ vs. PipeZK's 15 mm$^2$ NTT). We achieve MSM speedups of nearly 2$\times$ at large workload lengths. Our MSM runtimes include bucket reductions and window reductions performed on-chip, while PipeZK offloads reductions to CPU. We also achieve 3-8$\times$ NTT speedups. Note that for variable-length architectures, PipeZK's NTT suffers from low butterfly utilization. PipeZK uses a serially-pipelined NTT \cite{garrido} which uses 1 butterfly per NTT stage. PipeZK's NTT handles up to $2^{10}=1024$ elements to support a Four-step NTT of maximum length $N=2^{20}$. However, for workloads of length $N\leq2^{19}$, where $\sqrt{N} \leq 1024$, one or more stages of the NTT need to be bypassed during the row and/or column step; butterflies in those stages are then underutilized. In other words, except for workloads of maximal supported length, there are always some butterflies underutilized in \textit{each} PE of PipeZK's NTT. This poses a challenge to scalability because supporting longer NTTs risks more unused butterflies on sub-maximal-length workloads. In contrast, SZKP's memory-based NTT uses \textit{all} available butterflies in \textit{each} stage ensuring near-perfect utilization during operation.

\subsubsection{SZKP vs. GPU designs}
\label{szpk_v_gpu}

Table \ref{tab:combined_speedup_gpu} shows our speedups over cuZK and GZKP, which both use NVIDIA V100s, using 815 mm$^2$ in 12nm \cite{cuzk, gzkp, nvidia_v100_specs}. For data that both cuZK and GZKP provide, we show speedup over the faster design. We pick Dense MSM and NTT modules based on a  roughly $800$ mm$^2$ budget in 22nm. We use scale factors of 2$\times$ for area and 1.65$\times$ for delay \cite{28nmto16nm} to scale to 12nm. We pick an MSM with 8 PEs, 6-bit windows, and 1024 points per window (278 mm$^2$ in 12nm), and an NTT with 2 PEs and 4 butterflies/PE (20 mm$^2$ in 12nm), with Sparse G1 and G2 combined area of 102 mm$^2$. We choose this design to show that even with $3\times$ and $40\times$ less area than a V100 (and at 300 MHz), we still achieve $15-35\times$ MSM speedup and $2.5-4\times$ NTT speedup, respectively.

\begin{table}[ht]
\centering
\caption{Runtime (ms) of kernels vs CPU on 254b}
\vspace{-.1in}
\label{tab:combined_speedup}
\begin{tabular}{|c|c|c|c|c|}
\hline
\multirow{2}{*}{\textbf{Size}} & \multicolumn{2}{c|}{\textbf{CPU}} & \multicolumn{2}{c|}{\textbf{SZKP-7nm}} \\ \cline{2-5}
 & \textbf{Poly} & \textbf{MSM} & \textbf{Poly} &  \textbf{MSM} \\ \hline \hline
\raisebox{-1pt}{$\mathbf{2^{14}}$} & 102 & 344   & 0.078 (\textbf{1315$\times$})   & 0.52 (\textbf{662$\times$}) \\ \hline
\raisebox{-1pt}{$\mathbf{2^{15}}$} & 211 & 625   & 0.137 (\textbf{1545$\times$})   & 0.78 (\textbf{804$\times$}) \\ \hline
\raisebox{-1pt}{$\mathbf{2^{16}}$} & 479 & 1131  & 0.235 (\textbf{2040$\times$})   & 1.30 (\textbf{873$\times$}) \\ \hline
\raisebox{-1pt}{$\mathbf{2^{17}}$} & 982 & 2204  & 0.442 (\textbf{2223$\times$})   & 2.33 (\textbf{946$\times$}) \\ \hline
\raisebox{-1pt}{$\mathbf{2^{18}}$} & 2151  & 3884 & 0.807  (\textbf{2665$\times$})  & 4.40  (\textbf{883$\times$})\\ \hline
\raisebox{-1pt}{$\mathbf{2^{19}}$} & 4379  & 7554 & 1.567  (\textbf{2795$\times$})  & 8.53  (\textbf{885$\times$})\\ \hline
\raisebox{-1pt}{$\mathbf{2^{20}}$} & 8865  & 14834& 2.996  (\textbf{2959$\times$})  & 16.81 (\textbf{883$\times$})\\ \hline
\end{tabular}
\vspace{-4mm}
\end{table}

\begin{table}[ht]
\centering
\caption{Runtime (ms) of kernels vs PipeZK on 254b}
\vspace{-.1in}
\label{tab:combined_speedup_pipezk}
\begin{tabular}{|c|c|c|c|c|}
\hline
\multirow{2}{*}{\textbf{Size}} & \multicolumn{2}{c|}{\textbf{PipeZK}} & \multicolumn{2}{c|}{\textbf{SZKP}} \\ \cline{2-5}
 & \textbf{NTT} & \textbf{MSM} & \textbf{NTT} &  \textbf{MSM} \\ \hline \hline
\raisebox{-1pt}{$\mathbf{2^{14}}$} & 0.076 & 1  & 0.026  (\textbf{2.92$\times$})   & 0.93  (\textbf{1.07$\times$}) \\ \hline
\raisebox{-1pt}{$\mathbf{2^{15}}$} & 0.151 & 2  & 0.048  (\textbf{3.15$\times$})   & 1.42  (\textbf{1.41$\times$}) \\ \hline
\raisebox{-1pt}{$\mathbf{2^{16}}$} & 0.281 & 4  & 0.087  (\textbf{3.23$\times$})   & 2.40  (\textbf{1.66$\times$}) \\ \hline
\raisebox{-1pt}{$\mathbf{2^{17}}$} & 0.604 & 8  & 0.166  (\textbf{3.64$\times$})   & 4.37  (\textbf{1.83$\times$}) \\ \hline
\raisebox{-1pt}{$\mathbf{2^{18}}$} & 1.489 & 16 & 0.318  (\textbf{4.68$\times$})   & 8.30  (\textbf{1.93$\times$})\\ \hline
\raisebox{-1pt}{$\mathbf{2^{19}}$} & 4.052 & 32 & 0.631  (\textbf{6.42$\times$})   & 16.17 (\textbf{1.98$\times$})\\ \hline
\raisebox{-1pt}{$\mathbf{2^{20}}$} & 11    & 61 & 1.25  (\textbf{8.80$\times$})  & 31.90 (\textbf{1.91$\times$})\\ \hline
\end{tabular}
\vspace{-2mm}
\end{table}

\begin{table}[ht]
\centering
\vspace{-2mm}
\caption{Runtime (ms) of kernels vs GPU on 753b}
\hspace{-.15in}
\label{tab:combined_speedup_gpu}
\begin{tabular}{|c|c|c|c|c|c|}
\hline
\multirow{2}{*}{\textbf{Size}} & {\textbf{cuZK}} & \multicolumn{2}{c|}{\textbf{GZKP}} & \multicolumn{2}{c|}{\textbf{SZKP-12nm}} \\ \cline{2-6}
 & \textbf{MSM} & \textbf{NTT} & \textbf{MSM} & \textbf{NTT} &  \textbf{MSM} \\ \hline \hline
\raisebox{-1pt}{$\mathbf{2^{14}}$} & --   &  0.15  & 20    &  0.04   (\textbf{4.1$\times$})  & 0.86     (\textbf{23.3$\times$}) \\ \hline
\raisebox{-1pt}{$\mathbf{2^{16}}$} & --   &  0.49  & 50    &  0.18   (\textbf{2.8$\times$})  & 2.83     (\textbf{17.7$\times$}) \\ \hline
\raisebox{-1pt}{$\mathbf{2^{18}}$} & --   &  1.91  & 160   &  0.74   (\textbf{2.6$\times$})  & 10.69    (\textbf{15.0$\times$}) \\ \hline
\raisebox{-1pt}{$\mathbf{2^{19}}$} & 732  &    --  & --    &  1.49                    & 21.18    (\textbf{34.6$\times$}) \\ \hline
\raisebox{-1pt}{$\mathbf{2^{20}}$} & 1163 &  7.46  & 600   &  2.99   (\textbf{2.5$\times$})  &  42.15   (\textbf{14.2$\times$}) \\ \hline
\raisebox{-1pt}{$\mathbf{2^{21}}$} & 1960 &    --  & --    &  5.98                    &  84.09   (\textbf{23.3$\times$}) \\ \hline
\raisebox{-1pt}{$\mathbf{2^{22}}$} & 3608 & 33.67  & 2660  &  11.97  (\textbf{2.8$\times$})  &  167.98  (\textbf{15.8$\times$}) \\ \hline
\raisebox{-1pt}{$\mathbf{2^{23}}$} & 6635 &    --  & --    &  23.96                   &  335.75  (\textbf{19.8$\times$}) \\ \hline
\raisebox{-1pt}{$\mathbf{2^{24}}$} & --   & 141.4  & 11300 &  47.92  (\textbf{3.0$\times$})  &  671.29  (\textbf{16.8$\times$}) \\ \hline
\raisebox{-1pt}{$\mathbf{2^{26}}$} & --   & 602.53 & 40700 &  191.73 (\textbf{3.1$\times$})  &  2684.56 (\textbf{15.2$\times$}) \\ \hline
\end{tabular}
\vspace{-3mm}
\end{table}

\begin{table*}[t]
\centering
\caption{Full Proof Runtime vs. CPU on BN128}
\vspace{-.1in}
\label{full_proof}
\begin{tabular}{|c|c|c|c|c|c|c|c|c|c|c|c|}
\hline
\multirow{2}{*}{\textbf{Workload}} & \multirow{2}{*}{\textbf{Size}} & \multicolumn{4}{c|}{\textbf{CPU Time (s)}} & \multicolumn{5}{c|}{\textbf{SZKP-7nm Time (ms)}} & \multirow{2}{*}{\textbf{Speedup}} \\ \cline{3-11} & & \textbf{Poly} & \textbf{Dense} & \textbf{SpG1\&G2} & \textbf{Total} & \textbf{Poly} & \textbf{Dense} & \textbf{SpG1} & \textbf{SpG2} & \textbf{Total} & \\ \hline \hline
\textbf{AES}        & 16383    & 0.103  & 0.345   & 0.127  & 0.576   & 0.30   & 0.90  & 1.07   & 0.43  &  1.20    & \textbf{480$\times$} \\ \hline
\textbf{SHA2}       & 32767    & 0.212  & 0.626   & 0.051  & 0.890   & 0.51   & 1.54  & 2.05   & 0.03  &   2.05   & \textbf{434$\times$} \\ \hline
\textbf{RSA}        & 131071   & 0.994  & 2.207   & 0.520  & 3.723   & 2.11   & 4.73  & 3.84   & 0.95  &    6.85  & \textbf{544$\times$} \\ \hline
\textbf{RSASigVer}  & 131071   & 0.996  & 2.213   & 0.423  & 3.634   & 2.11   & 4.73  & 5.15   & 0.90  & 6.85     & \textbf{531$\times$} \\ \hline
\textbf{MerkleTree} & 131071   & 0.958  & 2.190   & 0.376  & 3.526   & 2.11   & 4.73  & 2.11   & 0.84  &   6.85   & \textbf{515$\times$} \\ \hline
\textbf{Auction}    & 1048575  & 8.912  & 14.834  & 0.919  & 24.666  & 18.63  & 34.44 & 36.87  & 1.02  &  53.07   & \textbf{465$\times$} \\ \hline
\end{tabular}
\vspace{-5pt}
\end{table*}

\begin{table*}[t]
\centering
\caption{Full Proof Runtime (ms) vs. PipeZK on MNT4753}
\vspace{-.1in}
\label{full_proof_pzk}
\begin{tabular}{|c|c|c|c|c|c|c|c|c|}
\hline
\multirow{2}{*}{\textbf{Workload}} & \multirow{2}{*}{\textbf{Size}} & \multicolumn{4}{c|}{\textbf{PipeZK}} & \multicolumn{3}{c|}{\textbf{SZKP}} \\ \cline{3-9} & & \textbf{Poly} & \textbf{G1 MSMs} & \textbf{Poly + G1} & \textbf{Total} &  \textbf{Poly} & \textbf{Dense} & \textbf{Total\tablefootnote{speedup over Poly+G1 / speedup over total}} \\ \hline \hline
\textbf{AES}      & 16383         & 2 &   21 &  23   & 97   &  1.03 \textbf{(1.94$\times$)} &  16.17 & 17.20 \textbf{(1.34$\times$ / 5.64$\times$)} \\ \hline
\textbf{SHA2}     & 32767         & 3 &   27 &  30   & 102  &  2.00 \textbf{(1.50$\times$)} &  31.08 & 33.09 \textbf{(0.91$\times$ / 3.08$\times$)} \\ \hline
\textbf{RSA}        & 98303/131071  & 14 &  80 &  94   & 1230 &  7.80 \textbf{(1.79$\times$)} &  90.28 & 98.09 \textbf{(0.96$\times$ / 12.54$\times$)} \\ \hline
\textbf{RSASigVer}  & 131071        & 14 &  105 & 119  & 822  &  7.80 \textbf{(1.79$\times$)} & 120.50 & 128.30 \textbf{(0.93$\times$ / 6.41$\times$)} \\ \hline
\textbf{MerkleTree} & 294911/524287 & 63 &  226 & 289  & 2697 & 32.01 \textbf{(1.97$\times$)} & 268.87 & 300.88 \textbf{(0.96$\times$ / 8.96$\times$)} \\ \hline
\textbf{Auction}    & 1048575       & 139 & 445 & 584  & 2053 & 65.39 \textbf{(2.13$\times$)} & 507.31 & 572.70 \textbf{(1.08$\times$ / 3.79$\times$)} \\ \hline
\end{tabular}
\vspace{-5pt}
\end{table*}

\begin{table*}[t]
\centering
\caption{Full Proof Runtime (ms) vs. GZKP on MNT4753}
\vspace{-.1in}
\label{full_proof_gpu}
\begin{tabular}{|c|c|c|c|c|c|c|c|}
\hline
\multirow{2}{*}{\textbf{Workload}} & \multirow{2}{*}{\textbf{Size}} & \multicolumn{3}{c|}{\textbf{GZKP}} & \multicolumn{3}{c|}{\textbf{SZKP-12nm}} \\ \cline{3-8} & & \textbf{Poly} & \textbf{MSM} & \textbf{Total} & \textbf{Poly} & \textbf{Dense} & \textbf{Total} \\ \hline \hline
\textbf{AES}      & 16383         & 4  &   99  &  103  &  0.31  \textbf{(12.8$\times$)} &  0.88  &  1.19 \textbf{(86.5$\times$)} \\ \hline
\textbf{SHA2}     & 32767         & 5  &   66  &  71  &  0.61   \textbf{(8.2$\times$)}&  1.54  &  2.15   \textbf{(33.0$\times$)} \\ \hline
\textbf{RSA}        & 98303/131071  & 22  &  120 &  142  &  2.37  \textbf{(9.3$\times$)} &  4.17  &  6.54  \textbf{(21.7$\times$)} \\ \hline
\textbf{RSASigVer}  & 131071        & 24  &  130 &  154  &  2.37  \textbf{(10.1$\times$)} &  5.56  &  7.93 \textbf{(19.4$\times$)} \\ \hline
\textbf{MerkleTree} & 294911/524287 & 60  &  220 &  280  &  9.71  \textbf{(6.2$\times$)} & 12.19  & 21.90  \textbf{(12.8$\times$)} \\ \hline
\textbf{Auction}    & 1048575       & 150  & 370 &  520  &  19.82 \textbf{(7.6$\times$)} & 22.89  & 42.71  \textbf{(12.2$\times$)} \\ \hline
\end{tabular}
\end{table*}

\vspace{-2mm}
\subsection{Full Chip Evaluations}

\subsubsection{Shared Vs. Separate G1}
\begin{figure}
  \centering
  \hspace{-5mm}
  \includegraphics[width=0.5\textwidth]{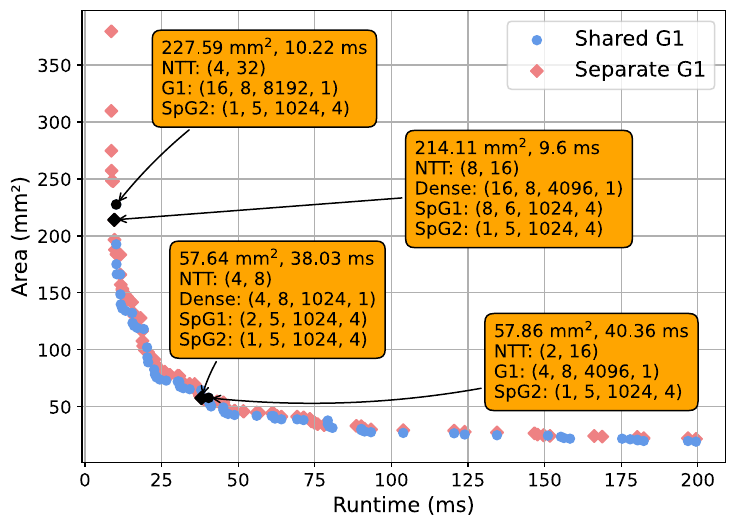}
  \vspace{-0.3in}
  \caption{Comparison of Shared-G1 vs. Separate-G1 Pareto-optimal designs for Auction. The NTT designs are denoted as ($K_N, U$), and MSMs are denoted as ($K_M, W, PPW, II$). For shared designs, we denote G1 for both Dense and Sparse MSM}
  \label{fig:shared_v_separate}
  \vspace{-3px}
\end{figure}

As mentioned in section \ref{sh_v_sp}, we can share the G1 MSM core between Sparse and Dense MSMs. We evaluate both topologies, Shared-G1 and Separate-G1, on Jsnark's Auction workload at 254b and plot Pareto curves in Figure \ref{fig:shared_v_separate}. We would typically expect separate (shared) hardware to dominate in high-performance, high-area (low-performance, low-area) configurations.

Surprisingly, from these plots we see that the shared topology does not yield significant performance benefits.
In fact, the shared-G1 topology's Pareto frontier essentially tracks the separate-G1 topology's frontier, except for high-performance (and high-area) where only Separate-G1 yields valid designs. The reason for this is that Sparse G1 MSMs are not on the critical path in low-performance designs; therefore, even though Separate-G1 requires extra PADDs, they are small and slow and add minimal area overhead. 

However, as we move to higher performance designs, Sparse G1 MSMs improve much slower than Dense MSMs and \emph{start appearing on the critical path}. Thus, having separately optimized Sparse G1 cores enables efficient resource allocation toward Dense MSMs.

\begin{figure}
  \centering
  \hspace{-5mm}
  \includegraphics[width=0.5\textwidth]{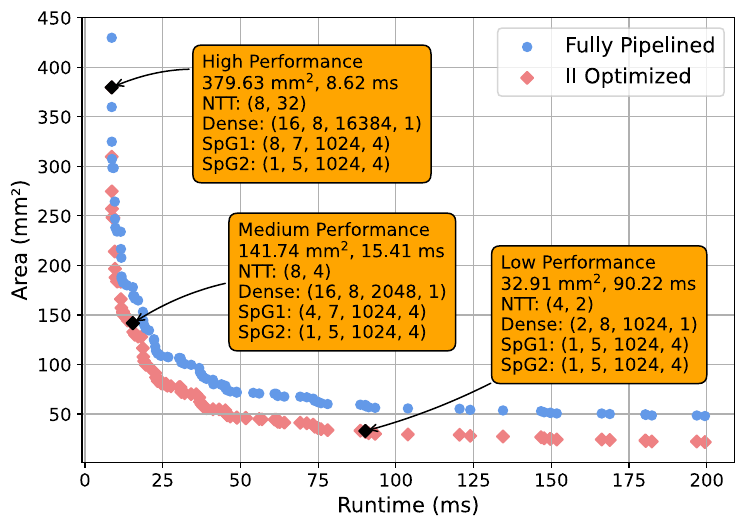}
  \vspace{-.24in}
  \caption{Effect of Pipelining on Area. HP, MP, and LP designs are highlighted.}
  \label{fig:ii_optimized}
\end{figure}
\begin{figure}
  \centering
  \hspace{-4mm}
  \includegraphics[width=0.5\textwidth]{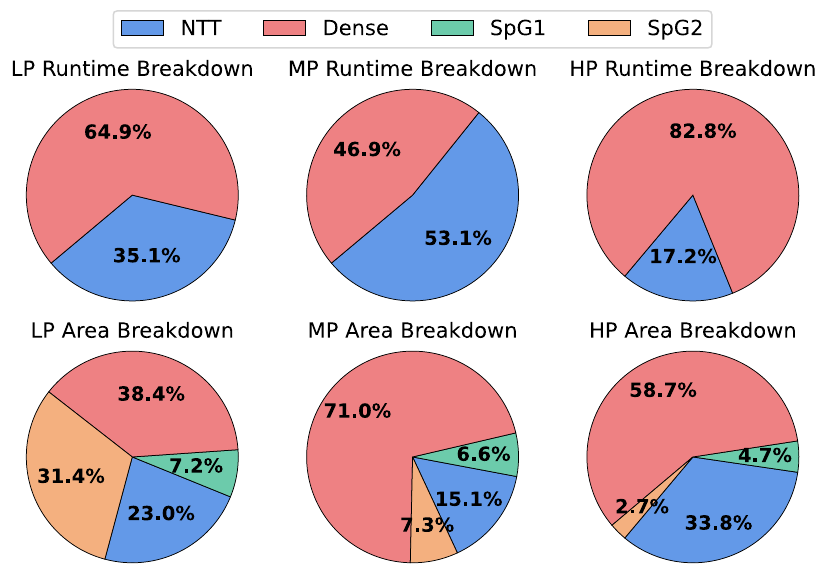}
  \vspace{-5mm}
  \caption{Runtime and Area breakdowns for chosen designs.}
  \label{fig:piecharts}
  \vspace{-10mm}
\end{figure}

\subsubsection{Optimizing Pipeline Depth}
How much does pipelining actually benefit Sparse MSMs? Figure \ref{fig:ii_optimized} compares the Pareto frontiers for 254b SZKP designs that have all Sparse MSMs fully pipelined versus II-optimized designs that save area. As we can see, there is virtually no performance penalty when we reduce pipelining. We highlight three designs, denoted as HP for high-performance, MP for medium-performance, and LP for low-performance. HP is chosen as the most performant and area intensive design, LP is chosen to be iso-area with our CPU (when scaled to 7nm), and MP is chosen as a rough midpoint along the Pareto curve. For each of these designs, we observe that \textit{both} Sparse G1 and G2 cores use an II = 4, demonstrating that they are not on the critical path. 

\subsubsection{Area and Runtime Breakdowns} Figure \ref{fig:piecharts} shows the runtime critical path and area breakdowns for each performance point. As expected, we see the Sparse G1/G2 PADD units progressively become smaller proportions of the area as Dense MSM and NTT cores become heftier. Interestingly, MP has a relatively balanced runtime. Designs similar to MP might be preferable for optimizing throughput at a \textit{proof}-level of pipelining. For example, MP designs could benefit \emph{proof-as-a-service} applications where a prover is computing multiple proofs for clients and is not significantly bottlenecked by the Dense MSM runtimes. This highlights yet another advantage of decoupling Sparse MSMs from the critical path, as it allows for designs that otherwise may not have appeared in the design space.

\subsubsection{Full-proof Speedup}

Similar to measuring the speedup of NTT/Poly and Dense MSMs, we measure the speedup of our architecture over CPU when scaled down to 7nm. We pick low-area designs near 33 mm$^2$ in 22nm and scale our runtimes to 7nm based on the aforementioned scale factors and report speedups in Table \ref{full_proof}. SZKP achieves speedups of 434-544$\times$ over CPU benchmarks.

Tables \ref{full_proof_pzk} and \ref{full_proof_gpu} show full-proof speedups over PipeZK and GZKP on 753b data. We show speedups over only Poly and Total runtimes because neither GZKP nor PipeZK parallelize Sparse MSMs from Dense MSMs. Furthermore, SZKP's NTT architecture only supports power-of-2 workload lengths, while PipeZK and GZKP report results for workloads that are of ``step-radix'' length and require a different polynomial algorithm \cite{libsnark} than shown in Figure \ref{fig:groth-dataflow}. For fair comparison, we report our power-of-2 length runtimes for Poly and  ``step-radix'' length runtimes for the Dense MSM, since MSMs are not bound to power-of-2 lengths.

For comparison with PipeZK, we pick an MSM with 1 PE, 5-bit windows, 1024 points per window (71 mm$^2$) and an NTT with 1 PE, 4 butterflies/PE (20 mm$^2$). These designs are dominated in area by memory due to double buffering used in both MSM and NTT modules; additionally, we use pessimistic memory area estimates due to the memory compiler's inability to directly generate 753b memories. Our total runtime is roughly as fast as PipeZK's excluding G2 computation. As previously mentioned, our runtime estimates include bucket and window reductions while PipeZK performs them on CPU; PipeZK does not report reduction latencies. Left un-accelerated, these reduction steps could dominate the overall MSM latency. With offloaded-G2 latencies included, we achieve $3-12$x speedup over PipeZK since G2 computations become their critical path. 
Additionally, PipeZK's 254b architecture is potentially less efficient than their 753b one. Given that their 753b single-PE NTT/MSM design is faster than G2-on-CPU, their 254b 4-PE design is likely much faster than G2-on-CPU. This implies more idle accelerator time while waiting for G2-on-CPU. Even if PipeZK accelerated G2, for all bit-widths, they would likely incur a high area footprint (that SZKP would not) from complex FIFO control logic, since G2 uses pairs of points. This further limits their scalability.

For comparison with GZKP, we use the same design mentioned in section \ref{szpk_v_gpu}. Adjusted for length, SZKP achieves $12-86$x full-proof speedup over GZKP, using \textit{half} the area with \textit{conservative} scaling factors at \textit{slow} ASIC frequencies. 

\vspace{-2mm}
\begin{table}[ht]
\centering
\caption{Full-Chip Power and Power Density}
\vspace{-3mm}
\label{tab:Thermal Design Power}
\begin{tabular}{|c|c|c|c|}
\hline
\multirow{2}{*}{\textbf{Design}} & \textbf{Area} & \textbf{TDP} & \textbf{Power Density} \\  & \textbf{(mm$^2$)} & \textbf{(W)} & \textbf{(W/mm$^2$)} \\ \hline \hline
\textbf{AMD EPYC 7502}  & 296   & 180 \cite{amd_specs}  & 0.61  \\ \hline 
\textbf{SZKP-7nm}       & 78.7  & 34   & 0.43 \\ \hline
\textbf{NVIDIA V100}    & 815   & 300 \cite{nvidia_v100_specs}  & 0.37 \\ \hline
\textbf{SZKP-12nm}      & 832   & 184  & 0.22  \\ \hline
\end{tabular}
\vspace{-2pt}
\end{table}

\vspace{-2mm}
\subsubsection{Power}

We estimate the thermal design power (TDP) for SZKP designs by constructing power traces of each module; we assume 100\% utilization to estimate worst-case, peak, full-chip power. We also include a 1 TB/s, 32W HBM stack \cite{bts, ark}. We look at high-performance designs for both 254b and 753b and select those with highest peak power. We then scale down the 254b design to 7nm (area by 3.6$\times$, power by 3.3$\times$) and the 753b design to 12nm (area by 2$\times$, power by 2.5$\times$) \cite{haac, 28nmto16nm}. We compare these two designs' TDP with that of the AMD EPYC 7502 (for 254b) and the NVIDIA V100 (for 753b). Table \ref{tab:Thermal Design Power} shows these comparisons. High-performance SZKP architectures fall well within the TDP and power density limits of high-performance CPUs and GPUs.

\subsubsection{Bandwidth Analysis}
Our initial Pareto analysis examines SZKP under \textit{unconstrained} bandwidth to identify peak performance across the design space. However, recent ZKP and privacy-centric accelerators have demonstrated high bandwidth needs. We extend our analysis by modeling SZKP under 6 memory technologies including DDR4 with 4 and 8 channels and four generations of HBM. 

\begin{figure*}[t!]
  \centering
  \includegraphics[width=0.97\textwidth]{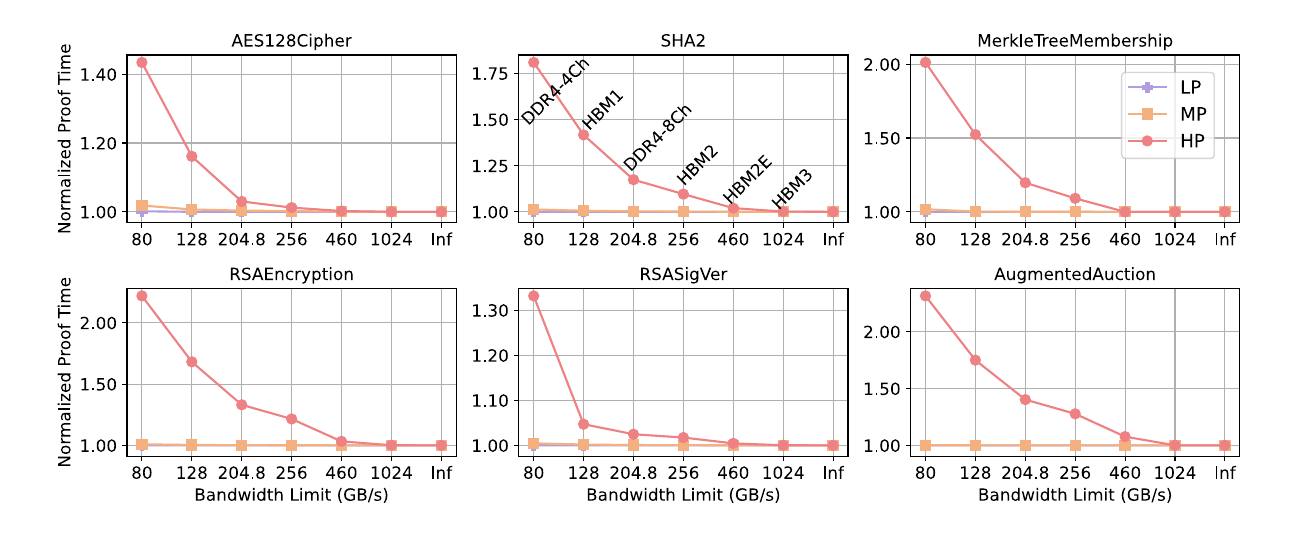}
  \vspace{-.22in}
  \caption{Normalized proof generation times on BN128 across memory technologies}
  \label{fig:bw_sweep}
  \vspace{-.16in}
  
\end{figure*}

Our results on 254b data are illustrated in Figure \ref{fig:bw_sweep}. We can see that across all workloads, low- and medium-performance design points can make do with a 4-channel DDR4 or HBM1. High-performance designs benefit from increased bandwidth, but here too HBM2E suffices to achieve close to ideal performance.

\begin{figure*}[t!]
  \centering
  \includegraphics[width=0.97\textwidth]{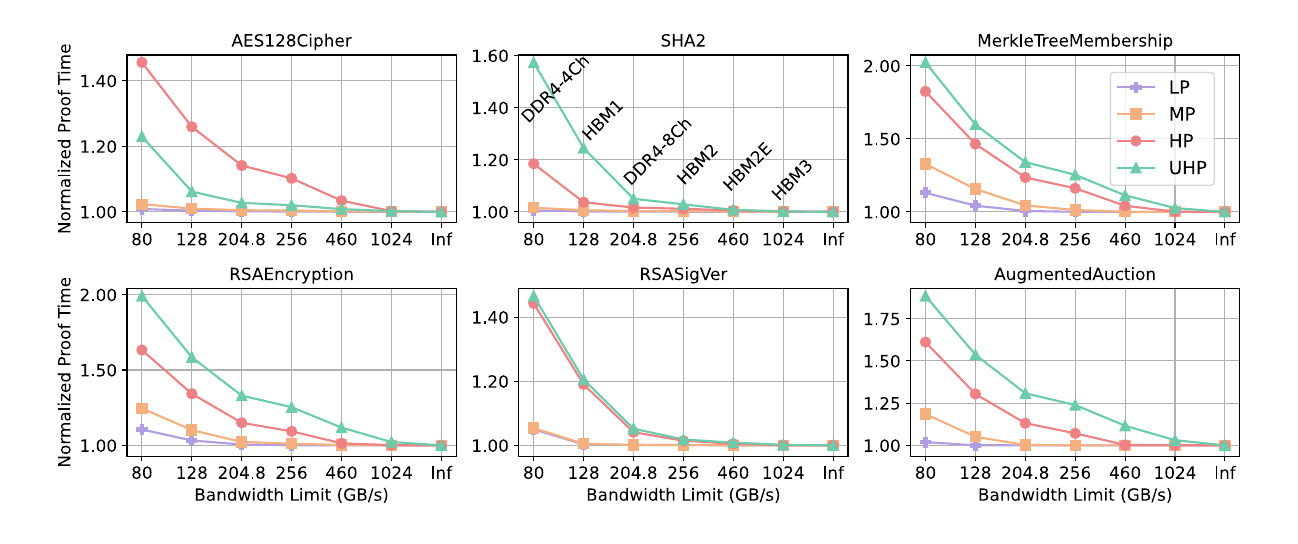}
  \vspace{-.22in}
  \caption{Normalized proof generation times on MNT4753 across memory technologies}
  \label{fig:bw_sweep_753}
  \vspace{-.1in}
  
\end{figure*}

 Extending our analysis to 753b, we choose 4 design points to reflect the wider range of areas in our design space. These are chosen to be iso-area (before scaling) with existing GPUs -- LP at 475 mm$^2$ to match an NVIDIA GTX 1080i \cite{tech_powerup_gtx}, MP at 815 mm$^2$ to match an NVIDIA Tesla V100 \cite{nvidia_v100_specs}, HP at 1600 mm$^2$ to match 2 V100s, and UHP (Ultra High Performance) at the largest design point per workload, ranging from 2000 mm$^2$ to 3500 mm$^2$ to match multiple V100s. UHP represents an extreme for ASICs, but we include it to highlight the range of architectures available in our optimal design space. These GPU references are based on those used by GZKP's implementation \cite{gzkp}. As seen in Figure \ref{fig:bw_sweep_753}, we find that even the large LP and MP designs can reach close to optimal performance with an 8-channel DDR4, and HP and UHP in several cases can make do with HBM2 or HBM2E, which are increasingly prevalent in recent fully homomorphic encryption accelerators \cite{f1, ark, bts, sharp, clake}. These results show that SZKP, when scaled to GPU sizes, still can perform optimally with comparatively modest bandwidth requirements.

\begin{figure}
  \centering
  \hspace{-5mm}
  \makebox[\columnwidth][c]{
    \includegraphics[width=0.49\textwidth]{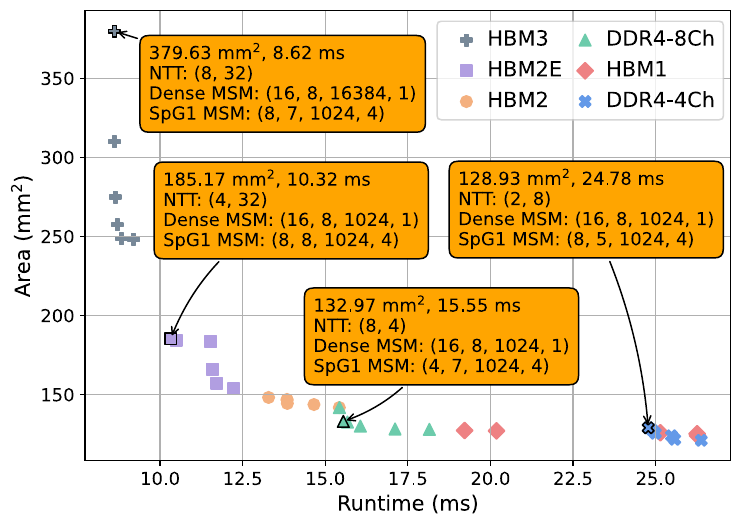}
  }
  \vspace{-.15in}
  \caption{Bandwidth-Aware Pareto Frontiers on BN128. All SpG2 modules have their parameters as (1, 5, 1024, 4).}
  \label{fig:bw_pareto}
  \vspace{-11pt}
\end{figure}

We next factor bandwidth constraints into the design space optimization. Figure \ref{fig:bw_pareto} shows the fastest Pareto-optimal designs under each bandwidth constraint on the longest workload (Auction) for 254b datatypes. Because the NTT unit is the bandwidth-intensive feature of SZKP, we see that resource-intensive NTTs are preferable under less-stringent bandwidth constraints. Notably, each of the highlighted design points has 16 PEs in the Dense MSM, indicating that SZKP's Dense MSM is clearly bandwidth efficient and scalable. We observe these trends across all workloads on both bit-widths.
These observations reinforce our intuitions that scalable ZKP design is achievable for ASICs when the correct dataflow is chosen.

\section{Related Work} {
PipeZK \cite{pipezk} is currently the only other custom hardware accelerator for full-proof zkSNARKs. PipeZK assumes relatively small (50 mm$^2$) chips and has difficulty scaling to larger designs because of the need for large multi-ported memories and complex control logic. SZKP, in contrast, is designed to easily scale from small to large designs, enabling a large design space to cover a range of applications. Additionally, for the MNT4753 curve, the Sparse G2 module has a significant contribution to overall chip area even after II optimization. This is because field multiplier (and PADD) sizes grow quadratically with field size, suggesting that full-chip evaluations including all computational modules are necessary to fully understand the benefits of custom ZKP accelerators.

GZKP \cite{gzkp} is a recent GPU-based ZKP hardware acceleration framework that leverages a custom finite field library to accelerate ZKP primitives. Their NTT module uses the traditional Cooley-Tukey NTT dataflow; as such, their NTT heavily depends on caching mechanisms to address the irregular memory access patterns. GZKP's MSM implementation is similar to SZKP in the sense that they perform bucket reductions after all points have been processed by a window, to avoid intermittent bucket reductions. Additionally, they eliminate window reductions, but they require  preprocessing overheads upwards of 5GB to store precomputed offsets for each window. They leverage a checkpointing scheme to achieve better balances between the space and time, but this solution, given the high memory overhead, is generally not scalable for ASIC-based designs. cuZK \cite{cuzk} is another recent GPU-based work that accelerates MSMs by converting all operations into sparse-matrix computations. They report linear speedup over Pippenger's algorithm, and additionally perform optimizations to minimize overheads for CPU-GPU data transfers. SZKP outperforms both of these prior works.

There is a large body of work on NTT acceleration, including hand-coded designs~\cite{f1, clake, bts, ark, sharp, pipezk} and designs generated using domain-specific languages like SPIRAL~\cite{spiral, spiral2}. SZKP uses HLS to synthesize high-performance NTT modules starting with a software implementation of constant-geometry NTTs. HLS is valuable because it enables both an easy design space exploration and a direct route from a software cryptography library to a hardware accelerator. As far as we know, prior work on HLS-generated NTTs only used Cooley-Tukey implementations which resulted in long synthesis runtimes and less performant designs~\cite{ncsu}. 

}

\vspace{-5pt}
\section{Conclusion} 
In this paper, we present SZKP, the \emph{first} ASIC framework that accelerates full zkSNARK proofs on-chip with a specific emphasis on scalability. Through a comprehensive design space exploration, we show how ZKPs can be constructed to yield orders of magnitudes of speedup over CPU, scale to GPU sizes while maintaining reasonable bandwidth needs, and outperform state-of-the-art GPUs and ASICs. SZKP makes a strong case for scalable ASIC solutions to the widespread adoption of ZKP hardware architectures.

\vspace{-2mm}
\begin{acks}
We thank the reviewers for their valuable comments. Support for this work was provided in part by NSF CAREER award \#2340137 and via a Google gift award, including GCP credits. This work was also supported by NSF RINGS \#2148293 and ARL.
\end{acks}

\bibliographystyle{ACM-Reference-Format}
\bibliography{ref}

\end{document}